\documentclass[12pt]{article}
\pdfoutput=1

\usepackage{array} 
\usepackage{amssymb}
\usepackage{graphics,graphpap}
\usepackage{graphicx}
\usepackage{color}
\usepackage{graphicx}
\usepackage{dcolumn}
\usepackage{epsfig}
\usepackage{epstopdf}
\usepackage{bbm}
\usepackage{amsmath}  
\usepackage{amsfonts}
\usepackage{textcomp}
\usepackage{cite}
\usepackage{setspace}

\newcommand{\beq}{\begin{eqnarray}}
\newcommand{\eeq}{\end{eqnarray}}

\newcommand{\bmp}{\noindent\begin{minipage}{16cm}}
\newcommand{\emp}{\end{minipage}\vskip 7mm} 

\def\drawbox#1#2{\hrule height#2pt
        \hbox{\vrule width#2pt height#1pt \kern#1pt
              \vrule width#2pt}
              \hrule height#2pt}

\def\Asym#1#2{\vcenter{\vbox{\drawbox{#1}{#2}
              \kern-#2pt 
              \drawbox{#1}{#2}}}}

\def\simge{\mathrel{%
   \rlap{\raise 0.511ex \hbox{$>$}}{\lower 0.511ex \hbox{$\sim$}}}}

\def\simle{\mathrel{
   \rlap{\raise 0.511ex \hbox{$<$}}{\lower 0.511ex \hbox{$\sim$}}}}

\def\s#1{\setbox0=\hbox{$#1$}%
\rlap{\ifdim\wd0>.7em\kern.22\wd0\else\kern.1\wd0\fi /}#1}

\begin{document}

\begin{titlepage}

\title{\vspace*{-2.0cm}
\hfill {\small UB-ECM-PF-87/13}\\
\vglue -0.3cm
\hfill {\small ICCUB-13-042} \vskip 0.2cm
\bf\Large
Influence of a keV sterile neutrino on neutrino-less double beta decay -- how things changed in the recent years }

\author{
Alexander Merle$^a$\thanks{email: \tt A.Merle@soton.ac.uk}~~~and~~Viviana Niro$^b$\thanks{email: \tt niro@ecm.ub.edu}
\\ \\
$^a${\normalsize \it Physics and Astronomy, University of Southampton,}\\
{\normalsize \it Southampton, SO17 1BJ, United Kingdom}\\
\\
$^b${\normalsize \it Departament d'Estructura i Constituents de la Mat\`eria and Institut de Ciencies del Cosmos,}\\
{\normalsize \it Universitat de Barcelona, Diagonal 647, E-08028 Barcelona, Spain}
}
\date{\today}
\maketitle
\thispagestyle{empty}

\vspace*{-1.0cm}
\begin{abstract}
\noindent
Earlier studies of the influence of Dark Matter keV sterile neutrinos on neutrino-less double beta decay concluded that there is no significant modification of the decay rate. These studies have focused only on a mass of the keV sterile neutrino above 2~and 4~keV, respectively, as motivated by certain production mechanisms. On the other hand, alternative production mechanisms have been proposed, which relax the lower limit for the mass, and new experimental data are available, too. For this reason, an updated study is timely and worthwhile. We focus on the most recent data, i.e., the newest Chandra and XMM-Newton observational bounds on the X-ray line originating from radiative keV sterile neutrino decay, as well as the new measurement of the previously unknown leptonic mixing angle $\theta_{13}$. While the previous works might have been a little short-sighted, the new observational bounds do indeed render any influences of keV sterile neutrinos on neutrino-less double beta decay small. This conclusion even 
holds in case not all the Dark Matter is made up of keV sterile neutrinos. 
\end{abstract}

\end{titlepage}

\section{\label{sec:intro}Introduction}

Sterile neutrinos with relatively small masses have been studied intensely within the recent years, due to both their phenomenological richness and due to experimental hints which could point towards their existence~\cite{Abazajian:2012ys}. One particularly interesting aspect of sterile neutrinos if they have a mass of a few keV is that they could potentially be the Dark Matter (DM)~\cite{Dodelson:1993je}, which makes up about $80\%$ of the matter content of the Universe~\cite{Komatsu:2010fb,Hinshaw:2012fq,Ade:2013lta}. A minimal framework for such neutrinos has been proposed in the form of the \emph{neutrino minimal standard model} ($\nu$MSM)~\cite{Asaka:2005an}, which is used to simultaneously accommodate for a variety of phenomena, such as neutrino oscillations, DM, or the baryon asymmetry of the Universe~\cite{Canetti:2012vf,Canetti:2012kh}. An important point is that keV neutrinos in such a framework typically have a \emph{warm} spectrum, i.e., they are neither highly relativistic (hot) DM, which would 
lead to problems with cosmological structure formation~\cite{Abazajian:2004zh,dePutter:2012sh}, nor are they non-relativistic (cold) DM. Extensive studies on neutrinos with masses of a few keV in the context of structure formation are present in the literature~\cite{Bode:2000gq,Hansen:2001zv,Boyarsky:2008xj,Lovell:2011rd,Boyanovsky:2010pw,Boyanovsky:2010sv,VillaescusaNavarro:2010qy}. In addition, surveys such as ALFALFA~\cite{Papastergis:2011xe} seem to point towards a DM mass of a few keV.

From the particle physics side, keV neutrinos are consistent with many frameworks, from variants of the scotogenic model~\cite{Sierra:2008wj,Gelmini:2009xd,Ma:2012if} to Left-Right symmetry~\cite{Bezrukov:2009th,Nemevsek:2012cd}. A rising field of research is the construction of mechanisms that can motivate the existence of the keV mass scale, see e.g.\ Refs.~\cite{Shaposhnikov:2006nn,Lindner:2010wr,Barry:2011fp,Barry:2011wb,Allison:2012qn,Heeck:2012bz,Araki:2011zg,Merle:2011yv,Kusenko:2010ik,Adulpravitchai:2011rq,Zhang:2011vh,Grossman:2010iq,Robinson:2012wu,Dias:2005yh,Cogollo:2009yi,Mavromatos:2012cc,Dev:2012bd}, or Ref.~\cite{Merle:2013gea} for a recent review. Ideally, these models should give predictions in combination with one of the known production mechanisms: while non-resonant production (Dodelson-Widrow (DW) mechanism~\cite{Dodelson:1993je}) is excluded for the case of zero lepton asymmetry~\cite{Seljak:2006qw,Palazzo:2007gz,Boyarsky:2008ju,Boyarsky:2008xj}, a large enough primordial asymmetry can 
lead to resonant non-thermal contributions (Shi-Fuller (SF) mechanism~\cite{Shi:1998km}), consistent with all bounds~\cite{Gorbunov:2008ka,Boyarsky:2008mt}. Further possibilities are the production via scalar decays~\cite{Shaposhnikov:1900zz} or the dilution of a thermal overproduction by entropy-producing decays of particles~\cite{Scherrer:1984fd}. These mechanisms have been applied, e.g., in Refs.~\cite{Bezrukov:2009th,Nemevsek:2012cd,Asaka:2006nq,Asaka:2006ek,Asaka:2006rw,Laine:2008pg,Wu:2009yr,Shaposhnikov:2006xi,Bezrukov:2012as,Khalil:2008kp,King:2012wg,Merle:2013wta}.

In particle physics, one of the most interesting questions is about the nature of neutrinos: are they \emph{Dirac} or \emph{Majorana} fermions? This question can, realistically, only be answered by \emph{neutrino-less double beta decay} ($0\nu\beta\beta$), a process where a nucleus decays to another one via the emission of only two electrons, $(A, Z) \to (A, Z+2) + e^- + e^-$, as recently reviewed in several references~\cite{Rodejohann:2011mu,Vergados:2012xy,Barabash:2011fg}. An observation of $0\nu\beta\beta$ would unambiguously prove that lepton number is violated~\cite{Schechter:1981bd,Hirsch:2006yk}, in contrast to the diagram-level prediction of the Standard Model. However, if this is the case, we still need more new physics besides $0\nu\beta\beta$ in order to generate a phenomenologically acceptable neutrino mass~\cite{Duerr:2011zd}. This observation opens up the possibility to investigate $0\nu\beta\beta$ in connection to neutrino mass models. Since many models for neutrino masses involve also 
sterile neutrinos, it is worthwhile to investigate $0\nu\beta\beta$ in this respect. Up to now, most investigations studied the contributions of very light ($\sim {\rm eV}$)~\cite{Barry:2011wb,Babu:2003is,Goswami:2005ng,Goswami:2007kv,Li:2011ss,Ghosh:2012pw} or relatively heavy ($\gg 100$~MeV)~\cite{Mitra:2011qr,Mitra:2012qz,Nemevsek:2011aa} sterile neutrinos. 

The influence of keV sterile neutrinos on the effective mass in $0\nu\beta\beta$ has also been discussed a few years ago in Ref.~\cite{Bezrukov:2005mx}, see also Refs.~\cite{Benes:2005hn,deGouvea:2006gz}. It was concluded that the restricted mass range, $2~{\rm keV} \lesssim M \lesssim 5~{\rm keV}$, in combination with the hard X-ray bound~\cite{Watson:2006qb,Abazajian:2001vt,Abazajian:2006jc,Boyarsky:2005us,Mirabal:2010jj} and the Ly-$\alpha$ bound~\cite{Boyarsky:2008xj}, renders the new contribution invisibly small. However, the caveat in that argument is two-fold. First, the lower bound from Ly-$\alpha$ is altered~\cite{Kusenko:2006rh} if an alternative production mechanism is considered~\cite{Bezrukov:2009th,Nemevsek:2012cd,Shaposhnikov:2006xi}. Second, in 2012 we have measured the previously unknown mixing angle $\theta_{13}$ to be relatively large~\cite{An:2012eh,Ahn:2012nd,Abe:2011fz,Abe:2012tg}: this, in turn, increases the variability of the effective mass~\cite{Lindner:2005kr}. Based on Ref.~\cite{
Bezrukov:2005mx}, the statement of a negligible influence of the keV sterile neutrino on $0\nu\beta\beta$ was repeated in Refs.~\cite{Bezrukov:2007qz,Shaposhnikov:2008rc,Boyarsky:2009ix,Ruchayskiy:2011aa,Asaka:2011pb}. In particular, the authors of Ref.~\cite{Asaka:2011pb} have applied an oversimplification by neglecting the CP-phase of the keV-neutrino contribution, which is however necessary to quantify its influence.

We will in the following explicitly calculate the influence of a keV sterile neutrino on $0\nu\beta\beta$ for comparatively light sterile neutrino masses. We thereby illustrate how our improved knowledge on the parameters involved has changed the picture within the last two years. Indeed, the arguments given in Refs.~\cite{Bezrukov:2005mx,Asaka:2011pb} disregarded exactly the part of the parameter space where a non-negligible influence of a DM keV sterile neutrino might have been present. However, in particular the X-ray bound -- which has been considerably improved recently -- destroys this possibly big influence. Thus, due to the new experimental results the conclusion of Refs.~\cite{Bezrukov:2005mx,Asaka:2011pb} remains correct after all, even if the low mass range below $2$~keV had been disregarded. 

Before starting our investigation, we want to point out that in the low mass range, below roughly 3~keV, there is another strong bound on the active-sterile mixing angle arising from the requirement that the sterile neutrinos produced by the DW-mechanism do not overclose the Universe. If taken at face value, this bound is even stronger than the X-ray bound~\cite{Canetti:2012vf,Canetti:2012kh}. However, the production of keV sterile neutrino DM is a very model-dependent point, and in general the early Universe is not perfectly understood. Even though the DW-production by the thermal plasma due to small admixtures to active neutrinos seems unavoidable, there are various mechanisms such as dilution by late entropy production, which could potentially wash out a DW-overproduction. The X-ray bound, on the other hand, is a very model-independent bound as long as active-sterile mixing is present at all in a certain setting. There might be further constraints from the mechanism generating the light neutrino mass: 
while, e.g., in a type~I seesaw setting~\cite{Minkowski:1977sc,Yanagida:1979as,GellMann:1980vs,Glashow:1979nm,Mohapatra:1979ia} the active-sterile mixing is relatively tighly constrained~\cite{Asaka:2011pb}, already a type~II~\cite{Magg:1980ut,Lazarides:1980nt} contribution would considerably change the situation (see, e.g., Ref.~\cite{Adulpravitchai:2011rq} for an illustrative example).

The viewpoint we are adapting in this paper is a phenomenological one, so we do not focus on one particular production mechanism and concentrate on the nearly model-independent X-ray bounds. However, since the DW-production is rather generic, in our figures we will mark also the regions obtained considering the DW bound at face value. Moreover, we provide an extended discussion of the applicability of our study to clarify in which situations our general conclusions hold and when additional more model-dependent bounds might be stronger. This should help the reader to decide whether or not potentially observable influences to $0\nu\beta\beta$ could appear in a concrete model.

The paper is organised as follows. In Sec.~\ref{sec:meff} we review the main expressions for the effective mass and present a detailed discussion of the bounds of the active-sterile neutrino mixing angle currently available. In Sec.~\ref{sec:strong} we discuss the possible influence of really light keV sterile neutrinos on the effective mass. We compare the situation before 2011 with the current one after the new Chandra and XMM-Newton observational results. A discussion of the applicability of our study and some related subtleties is given in Sec.~\ref{sec:applic}, before we summarize our results in Sec.~\ref{sec:conc}.

\section{\label{sec:meff}The effective mass}

The standard expression for the effective mass in neutrino-less double beta decay is given by (see, e.g., Refs.~\cite{Rodejohann:2011mu,Vergados:2012xy,Lindner:2005kr,Bilenky:1987ty})
\begin{equation}
 |m_{ee}^{(3)}| = |m_1 c_{12}^2 c_{13}^2 + m_2 s_{12}^2 c_{13}^2 e^{2i \alpha} + m_3 s_{13}^2 e^{2i \beta}|,
 \label{eq:mee_1}
\end{equation}
where $s_{ij} \equiv \sin \theta_{ij}$ and $c_{ij} \equiv \cos \theta_{ij}$ are functions of the mixing angles $\theta_{ij}$, and where $\alpha$ and $\beta$ are the Majorana phases. The superscript ``(3)'' refers to the fact that three generations of active neutrinos are contributing. If we now have one keV sterile neutrino in addition, where any other sterile neutrinos have masses much larger than the nuclear momentum transfer $|\vec{q}| = \mathcal{O}(100~{\rm MeV})$, the keV neutrino contribution will modify the above effective mass to~\cite{Barry:2011wb,Goswami:2005ng,Goswami:2007kv,Li:2011ss,Ghosh:2012pw,Blennow:2010th}
\begin{equation}
 |m_{ee}^{(4)}| \simeq |m_{ee}^{(3)} + M \theta^2 e^{2i \gamma}|,
 \label{eq:mee_2}
\end{equation}
where $M$ and $\theta$ are the mass and the active-sterile mixing angle of the keV neutrino. We have left it unspecified to which generation the keV neutrino belongs. The phase called $\gamma$ is actually a linear function of the fundamental Majorana phases in the full $6 \times 6$ neutrino mass matrix. Note that the authors of Ref.~\cite{Asaka:2011pb} have neglected the CP phase of the new keV sterile neutrino contribution, but this is however needed to fully quantify the influence on the effective mass. 

In the following, we will analyse Eq.~\eqref{eq:mee_2} for the two cases of normal (NO: $m_1 = m < m_2 = \sqrt{m^2 + \Delta m_\odot^2} < m_3 = \sqrt{m^2 + \Delta m_A^2}$) and inverted (IO: $m_3 = m < m_1 = \sqrt{m^2 + \Delta m_A^2} < m_2 = \sqrt{m^2 + \Delta m_\odot^2 + \Delta m_A^2}$) mass ordering of the light neutrinos, where $m$ is the smallest neutrino mass and $\Delta m_\odot^2 \equiv \Delta m_{21}^2$ ($\Delta m_A^2 \equiv |\Delta m_{31}^2|$) denotes the solar (atmospheric) mass square difference.

\begin{figure}[t]
\centering
\begin{tabular}{lr}
\includegraphics[width=8cm]{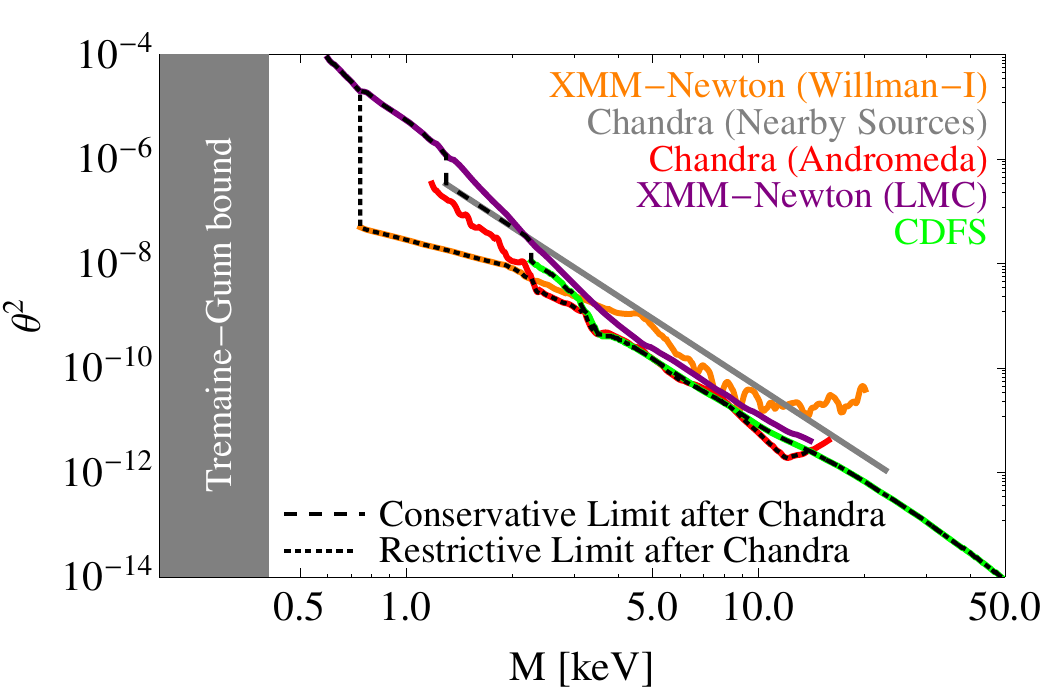} & \includegraphics[width=8cm]{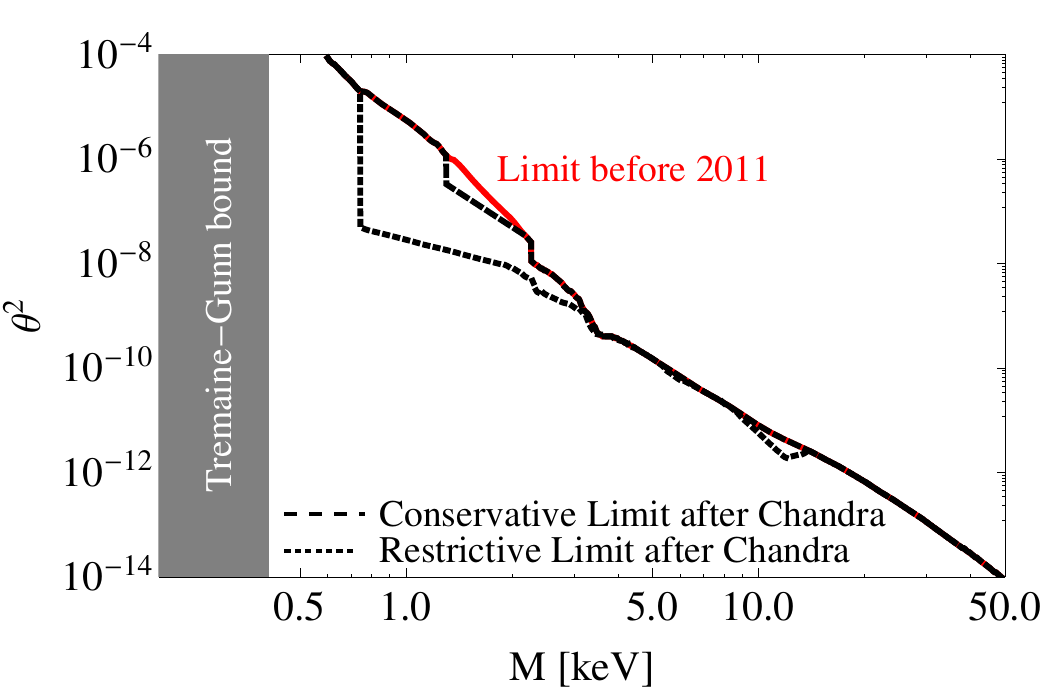}\\
\includegraphics[width=8cm]{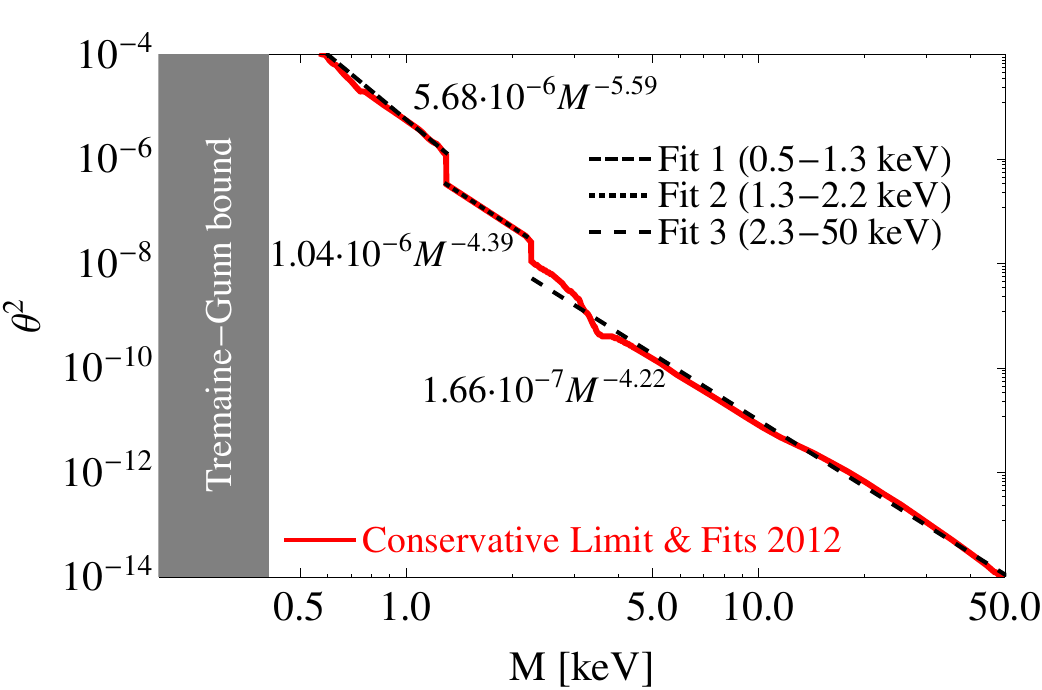} & \includegraphics[width=8cm]{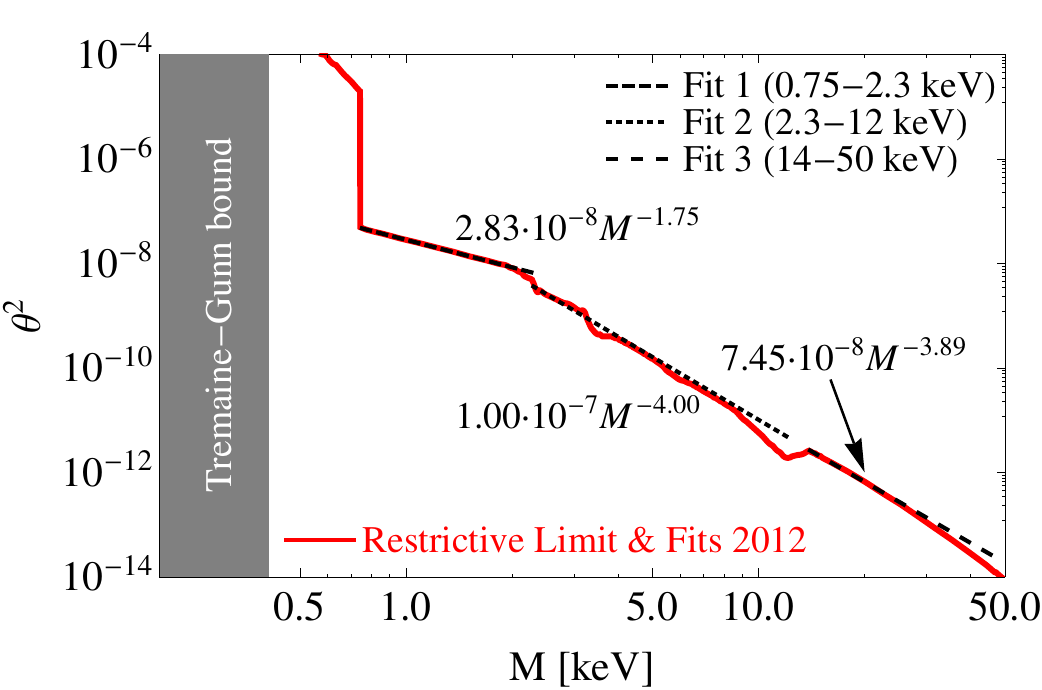}
\end{tabular}
\caption{\label{fig:bounds} Current and former X-ray bounds on the active-sterile mixing as functions of the keV sterile neutrino mass $M$, as well as the new bounds and suggested fits. Note that, within the fit formulas depicted in the lower two panels, $M$ is always taken to be measured in keV, but the unit is not shown for simplicity.}
\end{figure}

The important point to take into account is that the active-sterile mixing angle $\theta^2$ is strongly bounded by the non-observation of a monoenergetic astrophysical X-ray photon line stemming from the decay $N \to \nu \gamma$, where $N$ is the keV neutrino and $\nu$ is some light active neutrino. This bound cannot be avoided as long as active-sterile mixing is present (one could switch it off by artificially stabilizing the keV neutrino~\cite{Allison:2012qn}, but at the moment there is no convincing model known which predicts such a stabilization while at the same time giving a mechanism to generate the keV scale).\footnote{Note that stable keV neutrinos have been discussed in the context of the scotogenic model~\cite{Sierra:2008wj,Gelmini:2009xd,Ma:2012if}. However, this framework did not yield a mechanism to suppress the sterile neutrino mass scale.} The explicit observational bound, as valid before 2011, was summarized by Canetti, Drewes, Frossard, and Shaposhnikov (CDFS) in Refs.~\cite{Canetti:2012vf,Canetti:2012kh}
. This bound is based on the observations reported in Refs.~\cite{Watson:2006qb,Abazajian:2001vt,Abazajian:2006jc,Boyarsky:2005us,Dolgov:2000ew,Boyarsky:2006fg,RiemerSorensen:2006fh,Abazajian:2006yn,Boyarsky:2006ag,Boyarsky:2007ay,Boyarsky:2007ge,Loewenstein:2008yi}. 

However, our information on the mixing angle has changed considerably within the recent years. We have depicted the evolution using several example bounds in Fig.~\ref{fig:bounds}. In the upper left panel, the ``old'' observational bounds are depicted by the CDFS bound (green) and the XMM-Newton observations~\cite{Boyarsky:2006fg} of the Large Magellanic Cloud (LMC; purple), which goes down to slightly lower masses than CDFS.\footnote{Note that already in 2008, there has been a seemingly even stronger bound for very low masses originating from the Suzaku observations of Ursa Minor~\cite{Loewenstein:2008yi}. For larger masses, this data is included in the CDFS-fit~\cite{Canetti:2012vf,Canetti:2012kh}, but in particular for the small mass region the resulting bound seems to be stronger than the LMC limit~\cite{Boyarsky:2006fg}. However, it is highly non-trivial to compare the different data sets, as discussed in Sec.~3 of Ref.~\cite{Loewenstein:2008yi}, and there could be different scientific opinions on which 
limits to take into account, or not. We have, therefore, decided \emph{not} to include the Suzaku bound in our OLD scenario, in order to take on a conservative approach to the old limits. If the reader would like to recover the results including the Suzaku-limit, the bound would be very close to our scenario ``NEW restrictive'' to be discussed in the following, and its effect is therefore to some extent implicitly included in our plots.} In 2011, there had been new bounds by the results from the Chandra satellite~\cite{Watson:2011dw}, both for Nearby Sources (gray) and for the Andromeda galaxy (red). In addition, in 2012 this was extended by the bounds obtained by XMM-Newton from the observation of Willman~I~\cite{Loewenstein:2012px} (orange). Note that all the bounds shown in Fig.~\ref{fig:bounds} are at 99\% C.L. In particular, the bounds from Chandra satellite~\cite{Watson:2011dw}, both for Nearby Sources and Andromeda, have been rescaled from 95\% to 99\% and the bounds from XMM-Newton (LMC)~\cite{
Boyarsky:2006fg} and CDFS~\cite{Canetti:2012vf,Canetti:2012kh} have been rescaled from $3\sigma$ to 99\%. In addition, we have rescaled the CDFS bounds to be consistent with the estimated DM mass used by the other bounds, in particular the one by Chandra~\cite{Watson:2011dw}. Thus, the CDFS bounds reported in our figure are a factor two stronger than the one of Refs.~\cite{Canetti:2012vf,Canetti:2012kh}.\footnote{This bound had been artificially weakend by the same factor in Refs.~\cite{Canetti:2012vf,Canetti:2012kh} to account for possible uncertainties.}

However, we have not yet discussed how to combine these bounds. This is a subtle question, since after all different satellites have made observations of different galaxies, and when taking all observations at face value there could be unknown systematic errors involved. In order to show how such considerations can modify the results, we have decided to illustrate three different cases. This means we consider the limit as valid before 2011, consisting of the CDFS and the LMC bounds, of which we always take the strongest limit for a given mass. This limit is depicted as the red line in the upper right panel of Fig.~\ref{fig:bounds} and will later on be called ``OLD'' in the plots. We also use two different limits including the newer observations: the scenario ``NEW conservative'' adds only the Chandra observations of the nearby sources, while ``NEW restrictive'' also contains the Andromeda and Willman~I observations. These two scenarios are depicted in the upper right panel by the black dashed and dotted 
curves, respectively. We would like to stress that we make no judgement on which of the observations mentioned should be included into a more robust combined bound, or not. However, any possible choice is likely to yield results in between our two ``NEW'' scenarios. In the lower two panels of Fig.~\ref{fig:bounds}, we suggest linear fits in the $\log \theta^2$-$\log M [{\rm keV}]$ plane for our two scenarios. 

\begin{figure}[t]
\centering
\begin{tabular}{lr}
\includegraphics[width=8cm]{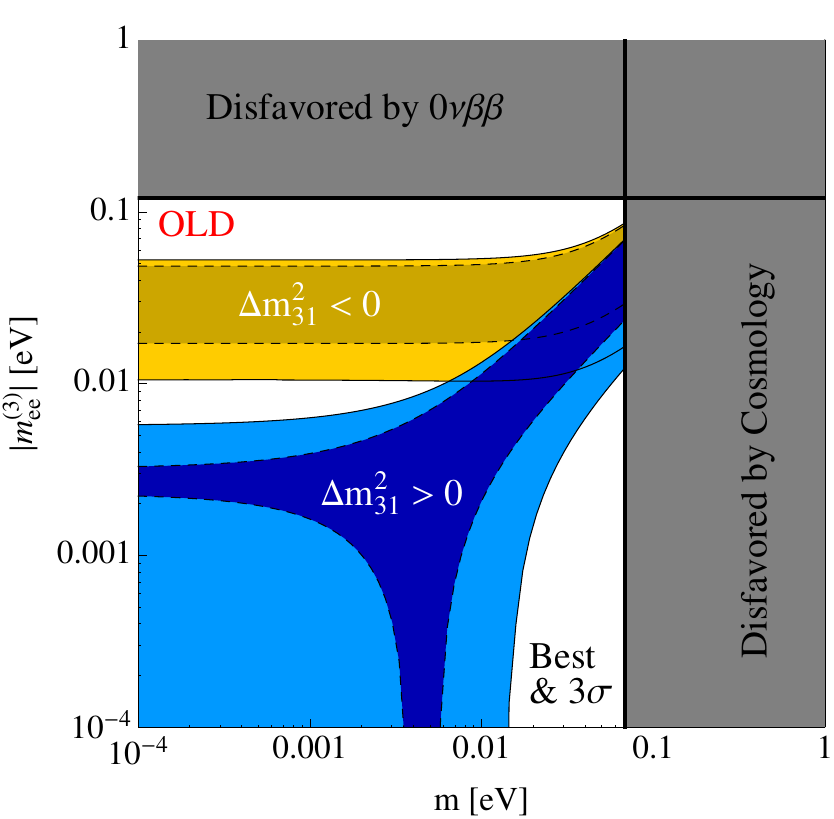} & \includegraphics[width=8cm]{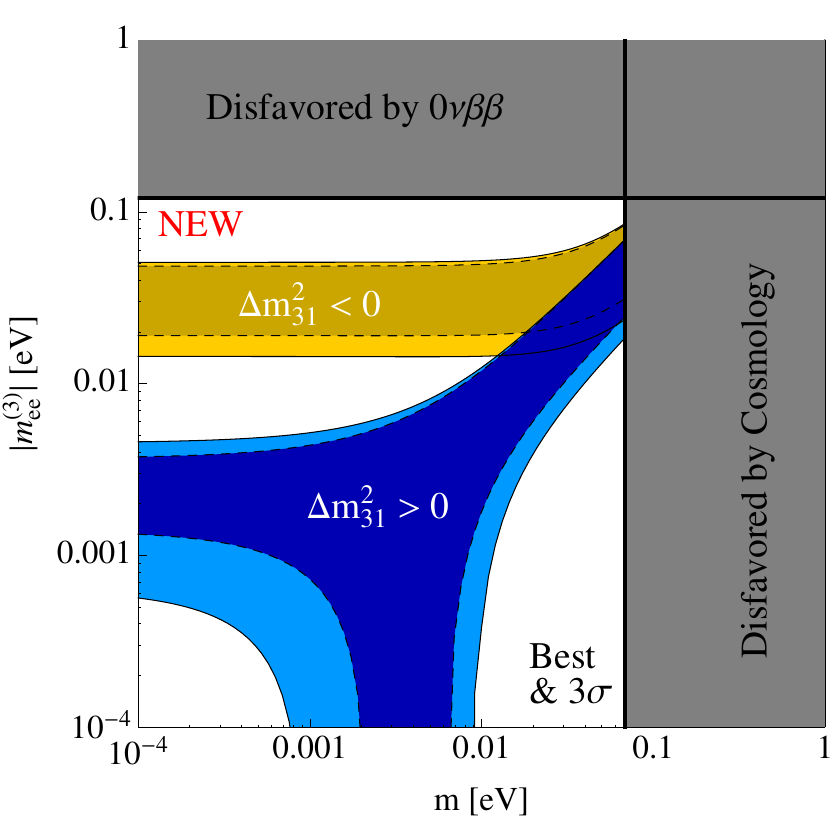}
\end{tabular}
\caption{\label{fig:mee} The change of the standard effective neutrino mass $|m_{ee}^{(3)}|$ in neutrino-less double beta decay with the new data.}
\end{figure}

An equally interesting development as for active-sterile mixing has taken place in the leptonic mixing data. In particular, the discovery of the non-zero value of the previously unknown leptonic mixing angle $\theta_{13}$ by the Daya Bay~\cite{An:2012eh}, RENO~\cite{Ahn:2012nd}, and Double Chooz~\cite{Abe:2011fz,Abe:2012tg} experiments has improved our knowledge. New global fits on the neutrino mixing data have recently appeared~\cite{GonzalezGarcia:2012sz,Tortola:2012te,Fogli:2012ua}. To give a flavour of the changes, we compare a recent global fit~\cite{GonzalezGarcia:2012sz} (free fluxes) on the new data with an older fit~\cite{GonzalezGarcia:2010er} (old Gallium fluxes) that had been updated in 2011. The decisive mixing parameters and their values (best-fit values and $3\sigma$ ranges) are:
\begin{center}
\begin{tabular}{|c||c|c|}\hline
Parameters & Old fit~\cite{GonzalezGarcia:2010er} & New fit~\cite{GonzalezGarcia:2012sz} \\ \hline \hline
$\sin^2 \theta_{12}$ & $0.32$ ($0.27$--$0.37$) & $0.30$ ($0.27$--$0.34$) \\ \hline
$\sin^2 \theta_{13}$ & $0.0095$ ($0.000$--$0.047$) & $0.023$ ($0.016$--$0.030$) \\ \hline
$\Delta m_\odot^2\ [10^{-5} {\rm eV}^2]$ & $7.59$ ($6.90$--$8.20$) & $7.50$ ($7.00$--$8.09$) \\ \hline
$|\Delta m_A^2|_{\rm NO}\ [10^{-3} {\rm eV}^2]$ & $2.46$ ($2.09$--$2.83$) & $2.47$ ($2.27$--$2.69$) \\ \hline
$|\Delta m_A^2|_{\rm IO}\ [10^{-3} {\rm eV}^2]$ & $2.36$ ($1.99$--$2.73$) & $2.43$ ($2.24$--$2.65$) \\ \hline
\end{tabular}
\end{center}
We will consider the old fit together with the X-ray limit before 2011, and the new fit with the conservative and restrictive limits after the Chandra and XMM-Newton results. 

The standard plot of the effective mass $|m_{ee}^{(3)}|$, cf.\ Eq.~\eqref{eq:mee_1}, is shown in Fig.~\ref{fig:mee}, while in the following sections we will present the influence of a keV-sterile neutrino contribution. Note that in the $0\nu\beta\beta$ plots, certain regions in the parameter space are disfavoured. From cosmology we obtain an upper limit on the sum $\Sigma$ of light neutrino masses, which in our setting (without sterile neutrinos at the eV scale) means that $\Sigma = m_1 + m_2 + m_3$. We take the upper limit $\Sigma < 0.230~{\rm eV}\ @95\%$~C.L.\ from the Planck 2013 data in combination with the WMAP polarization data, the data from the South Pole Telescope (SPT), the Atacama Cosmology Telescope (ACT), and from BAO (baryon acoustic oscillation), see Ref.~\cite{Ade:2013lta}. This limit can be translated into an upper limit on the lightest neutrino mass $m$, which is depicted in the plots. However, there could be unknown systematic errors involved, which are actually known to be able to lead to 
wrong conclusions about the absolute neutrino mass scale~\cite{Maneschg:2008sf}. This is why we mark that region as ``disfavoured'' rather than ``excluded''.

Similarly, for limits coming from $0\nu\beta\beta$-searches, there are always uncertainties from unknown nuclear physics~\cite{Rodejohann:2011mu}. Thus we take the corresponding parameter region to be ``disfavoured'' as well, and as example we take the most optimistic limit obtained on $|m_{ee}|$ from EXO-200~\cite{Auger:2012ar}. Both disfavoured regions are marked by the gray areas in Fig.~\ref{fig:mee}, Fig.~\ref{fig:keV_100}, and Fig.~\ref{fig:keV_50}.

\section{\label{sec:strong}Dependence for low masses}

The lowest possible keV neutrino mass is $M \simeq 0.4$~keV~\cite{Boyarsky:2008ju}. It results from the Tremaine-Gunn~\cite{Tremaine:1979we} bound, which originates from the fact that neutrinos are fermions. This bound can be regarded as more or less model-independent lower limit.\footnote{Note that a lower bound of $M \simeq 1$~keV was estimated in Ref.~\cite{Gorbunov:2008ka} based on similar assumptions.} If stronger Ly-$\alpha$ bounds are considered, the production mechanisms should be taken into account. For example, if the keV sterile neutrinos~\cite{Bezrukov:2009th,Nemevsek:2012cd,Asaka:2006nq,Asaka:2006ek} are produced by thermal overproduction with subsequent dilution by entropy production~\cite{Scherrer:1984fd}, the bound is relatively weak, $M \gtrsim 1.6$~keV~\cite{Bezrukov:2009th}, since the production of additional entropy leads to an effective cooling of the keV sterile neutrino component of the Universe~\cite{Abazajian:2012ys}. One could also build up part of the DM by resonant non-thermal 
production (Shi-Fuller mechanism~\cite{Shi:1998km}). This mechanism also leads to a cooler DM-spectrum, and hence the Ly-$\alpha$ bound is rescaled to a value of about $2$~keV~\cite{Boyarsky:2008mt}. Note that the resonant behaviour appears because in the case of large enough primordial lepton-antilepton asymmetries in the early Universe, Mikheev-Smirnov-Wolfenstein like~\cite{Wolfenstein:1977ue,Wolfenstein:1978ui,Wolfenstein:1979ni,Mikheev:1986gs,Mikheev:1986wj,Mikheev:1986if} level-crossings could appear. Similarly, the production by scalar decays at higher temperatures could lead to a diluted spectrum, and a lower bound on the keV neutrino mass would be relaxed by a factor of a few. Finally, the simplest production mechanism of keV sterile neutrinos -- the Dodelson-Widrow mechanism~\cite{Dodelson:1993je} -- in the framework of the $\nu$MSM produces a too warm DM spectrum, for which the lower Ly-$\alpha$ bound on the mass is between $8$~and $10$~keV~\cite{Boyarsky:2008mt}. This region is, in the case of 
vanishing primordial lepton asymmetry, already excluded by the X-ray bound~\cite{Canetti:2012vf,Canetti:2012kh}.

In general, the scenario proposed by Dodelson and Widrow~\cite{Dodelson:1993je}, although being plain and simple, has to be modified to obtain the correct DM relic abundance. Such modifications could decrease a potential overabundance produced by DW. A good example is the production of keV sterile neutrinos by thermal freeze-out, which is later on diluted by a release of entropy. The DM production of keV steriles is highly model-dependent, and therefore we cannot assume a very specific situation in a phenomenological study like the present one. However, we will try to at least cover the most extreme cases. In our analysis, we will mainly focus on the case in which the nearly model-independent X-ray bound is taken at face value\footnote{In this context, ``nearly model-independent'' simply means that the bound applies to models with active-sterile neutrino mixing. Note, however, that there are settings in which this mixing is \emph{not} present. One example are keV sterile neutrinos in the scotogenic model, 
see Refs.~\cite{Sierra:2008wj,Gelmini:2009xd}: the RH neutrino fields are charged non-trivially under a global $Z_2$ symmetry which is unbroken, while the LH neutrinos are singlets. In that case, there exists no mass term mixing active with sterile neutrinos, and hence the keV neutrino, if it is the lightest sterile neutrino, will be absolutely stable. Such settings are not constrained by the X-ray bound, but due to the absence of the mixing they also cannot lead to any non-trivial contributions to $0\nu\beta\beta$, which is why they are not relevant for the study presented here.} and on the case in which the upper bound on the active-sterile mixing angle is such that the keV sterile neutrino abundance, obtained through the DW mechanism, does not exceed the currently allowed $3\sigma$ value~\cite{Ade:2013lta}. To obtain a simple estimate of this bound, we have used the analytical approximation for the DM abundance given in Ref.~\cite{Kusenko:2009up}.

\begin{figure}[t]
\centering
\begin{tabular}{lr}
\includegraphics[width=8cm]{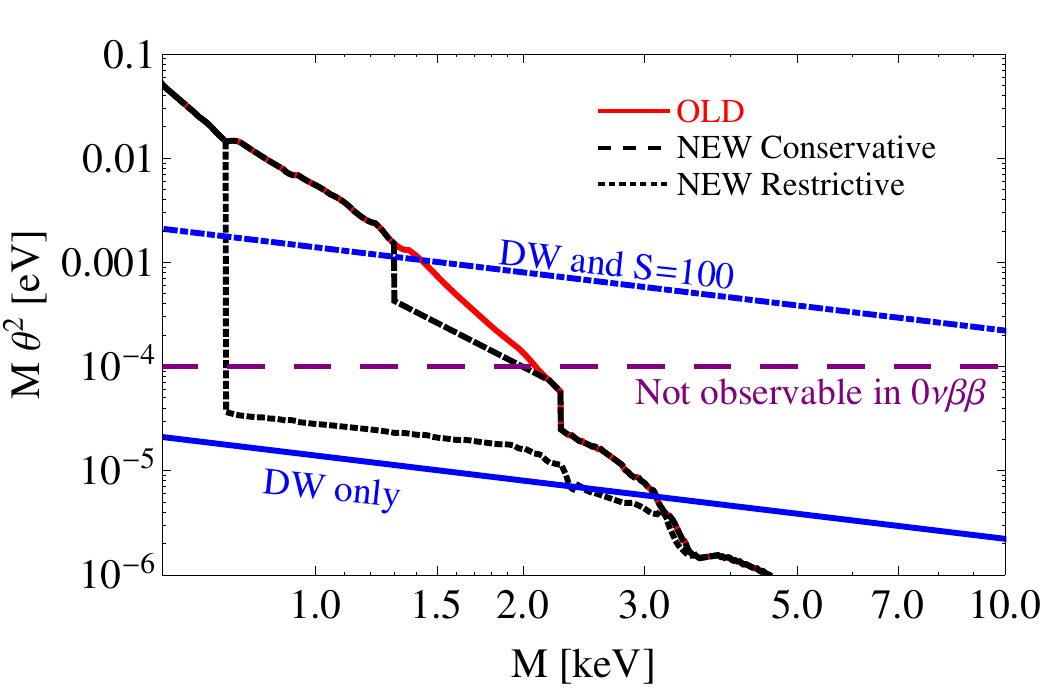} & \includegraphics[width=8cm]{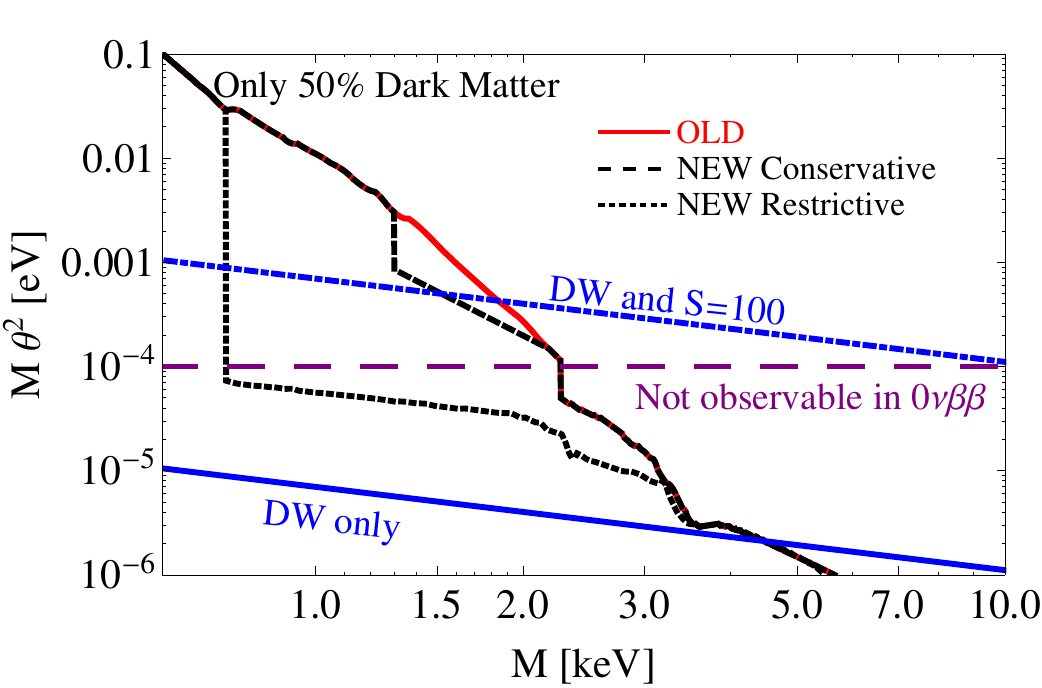} 
\end{tabular}
\caption{\label{fig:Bounds_sterile} Sterile neutrino contribution $M \theta^2$ to the $0\nu\beta\beta$ effective mass as a function of the sterile neutrino mass $M$. We present the results considering the OLD limits on the active-sterile mixing angle from X-ray bounds and the NEW limits, analysed with a restrictive and a conservative approach. We also show the limits on $M \theta^2$ requiring that the DM is not overproduced by the DW-mechanism only or by the DW-mechanism combined with an entropy dilution factor $S=100$. In the left panel we show the case in which the sterile neutrino represents all the DM present in the Universe, while in the right panel we show the case in which only 50\% of the DM is in the form of sterile neutrinos. Note that, in the latter case, while the X-ray bound gets weaker by a factor of roughly two (since less keV neutrinos exist in the Universe), the DW-bound gets \emph{stronger} by the same factor, since only a smaller amount of DM is allowed to be produced.}
\end{figure}

To first of all quantify the size of the keV sterile neutrino contribution to the effective mass $|m_{ee}^{(4)}|$, we show in Fig.~\ref{fig:Bounds_sterile} its absolute value $M\theta^2$, for the case in which the sterile neutrino constitutes all the DM present in the Universe (left panel) or only 50\% (right panel). It is well visible from the plots that a non-negligible contribution can be obtained for $M\lesssim2$~keV in case the OLD X-ray bounds or the NEW conservative bounds are considered. Using the NEW restrictive bounds, instead, this contribution is negligible and not observable in $0\nu\beta\beta$. However, as stated before, the DW contribution is hardly avoidable. Considering the bound on the mixing angle $\theta$, arising from not exceeding the correct relic abundance through the DW production mechanism, the contribution of the keV sterile neutrinos to $0\nu\beta\beta$ becomes too small to be relevant for $0\nu\beta\beta$ experiments, as indicated by the solid blue line in the plots. Despite this,
 the DW-bound can be considerably weakened in certain settings, for example, considering a subsequent entropy production through particle decays. Taking an entropy production of the order $\mathcal{O}(100)$ (cf.\ dashed blue line in the plots), the contribution of keV sterile neutrinos with $M\lesssim$~2~keV can be sizable and, at least in principle, observable in $0\nu\beta\beta$. Finally, note that in the full contribution of keV sterile neutrinos to the effective mass $|m_{ee}^{(4)}|$ a phase is present, too, see Eq.~\ref{eq:mee_2}. For this reason, if one wishes to quantify in detail the effect of the keV sterile neutrinos to $|m_{ee}^{(4)}|$, the study of the absolute value $M\theta^2$ is not sufficient, but a full analysis of $|m_{ee}^{(4)}|$ is required.

\begin{figure}[ht!]
\hspace{-1.0cm}
\begin{tabular}{lcr}
\includegraphics[width=5.5cm]{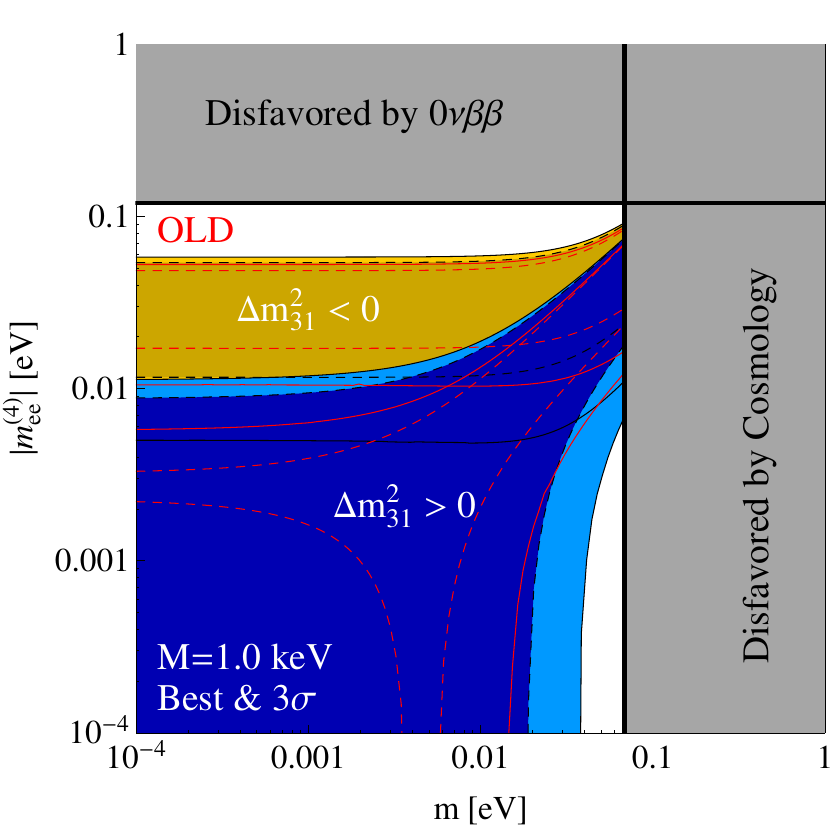} & \includegraphics[width=5.5cm]{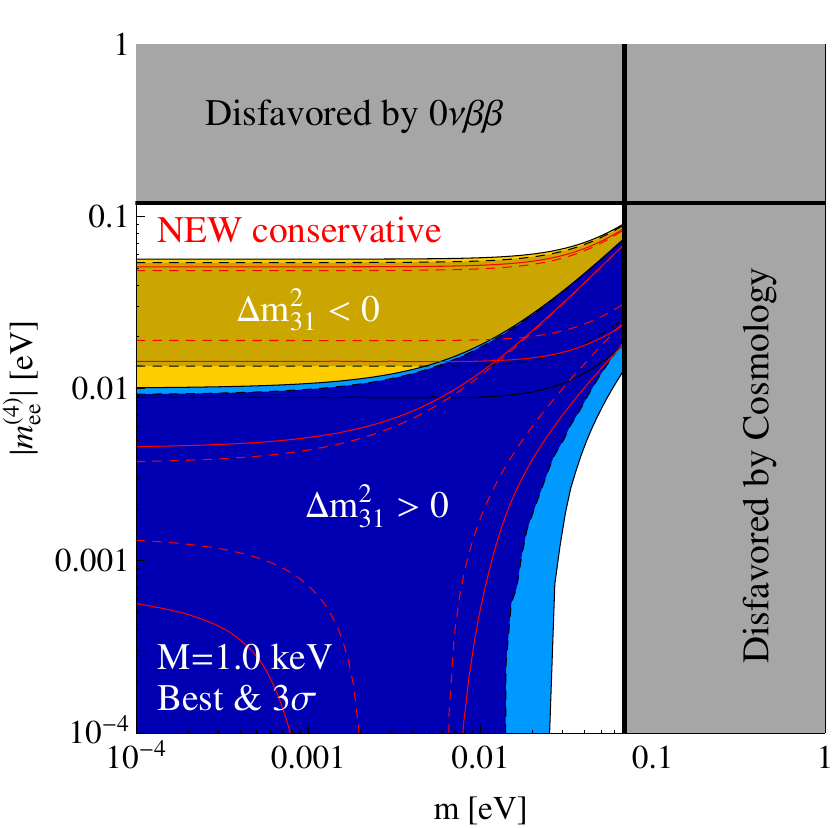} & \includegraphics[width=5.5cm]{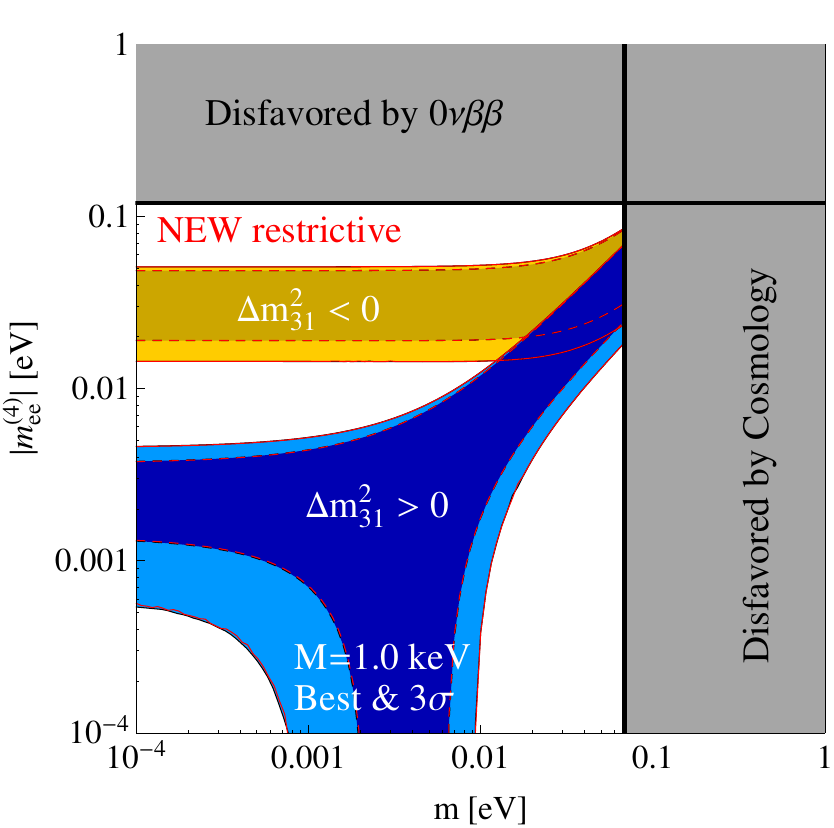}\\
\includegraphics[width=5.5cm]{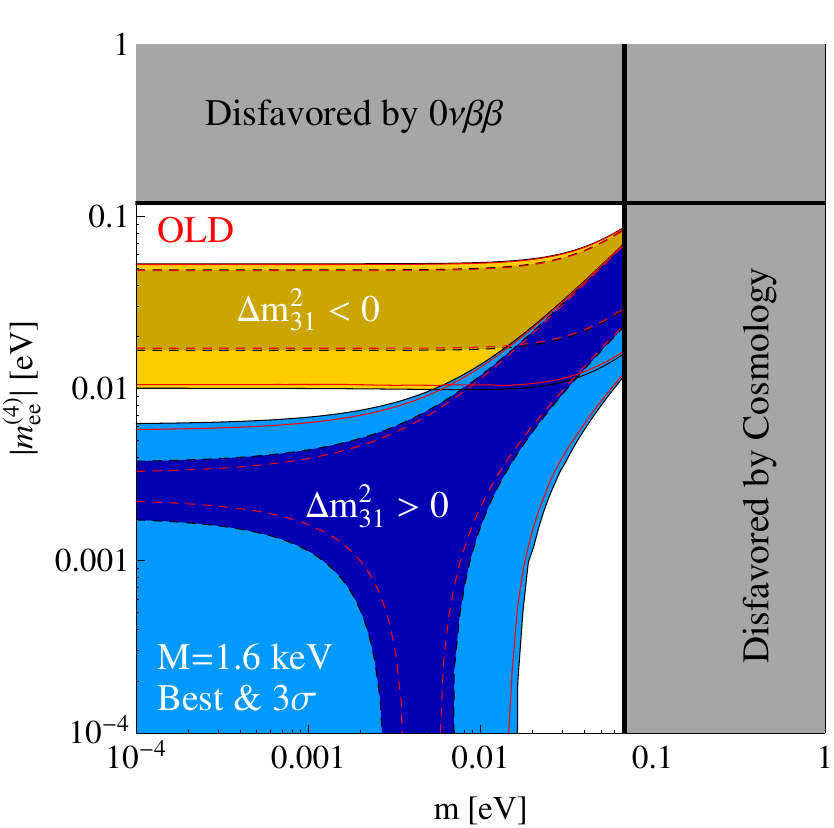} & \includegraphics[width=5.5cm]{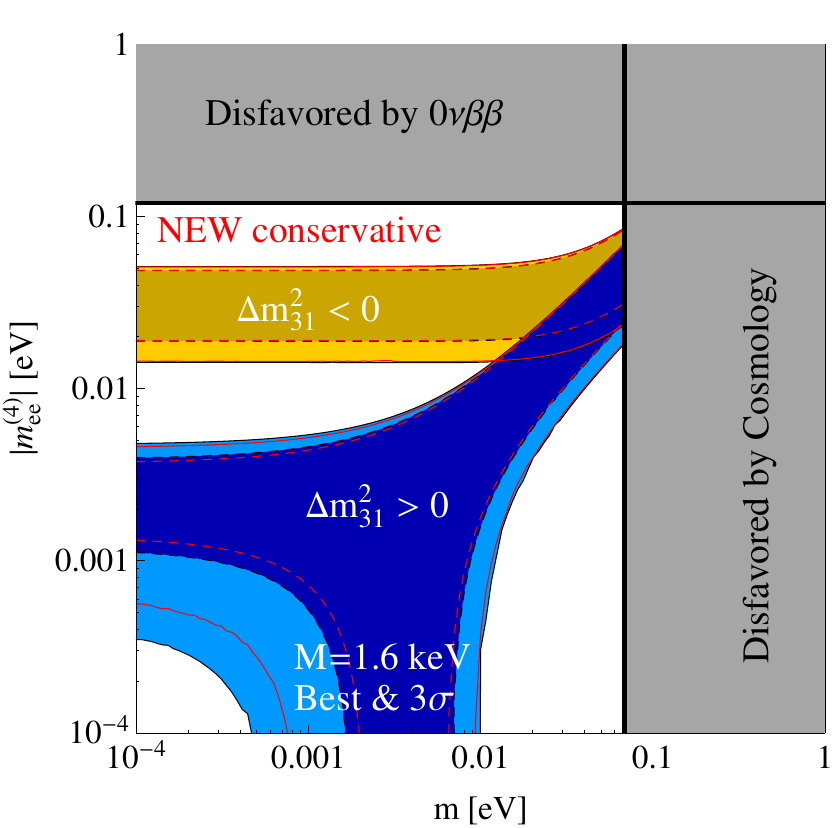} & \includegraphics[width=5.5cm]{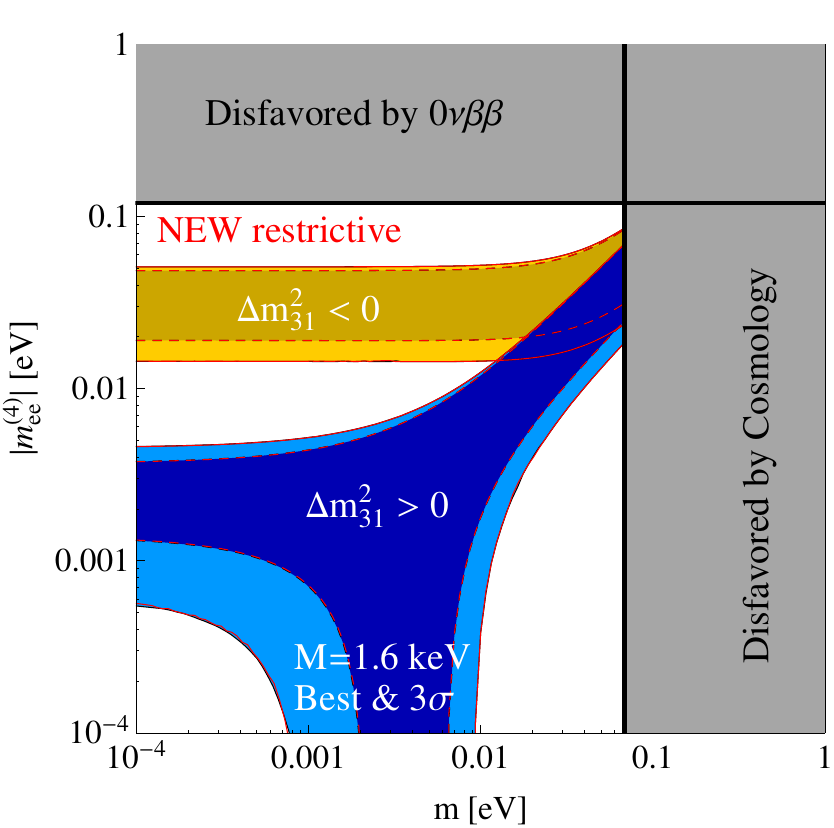}
\end{tabular}
\caption{\label{fig:keV_100} Effective mass $|m^{(4)}_{ee}|$ as a function of the smallest neutrino mass $m$. The left columns refer to the X-ray bounds before 2011 and to the old neutrino fits for mixing and mass parameters. The middle (right) columns refer to the updated conservative (restrictive) Chandra bounds and to the new neutrino fits. We refer to Sec.~\ref{sec:strong} for more details. The plots assume that the sterile neutrinos constitute all the DM present in the Universe. The thin red lines are obtained if the additional requirement of not overproducing the DM through the DW-mechanism is imposed.}
\end{figure}

A few example plots of $|m_{ee}^{(4)}|$ for two different values of the mass of the keV sterile neutrino are displayed in Fig.~\ref{fig:keV_100}. In the left column we vary the neutrino oscillation parameters within the corresponding old $3\sigma$ ranges~\cite{GonzalezGarcia:2010er} and the active-sterile mixing angle $\theta^2$ between zero and the OLD X-ray bounds as valid before 2011. In the middle and right columns, we display the results considering the limits derived from the NEW Chandra and XMM-Newton results: the conservative and the restricted bounds, respectively. For these cases, we use the newer best-fit values and $3\sigma$ ranges of Ref.~\cite{GonzalezGarcia:2012sz}. We consider two different values of the sterile neutrino mass: $M=1, 1.6$~keV (upper and lower rows, respectively). As illustrated in Fig.~\ref{fig:Bounds_sterile}, larger DM masses do not lead to a significant contribution to $0\nu\beta\beta$. In other settings, however, one might also have to take into account the bound arising 
from the necessity of not producing too much DM by the DW-mechanism, in which case the black solid (black dashed) lines would be altered to the red solid (red dashed) ones.

Let us first consider the situation valid before 2011, cf.\ left column of Fig.~\ref{fig:keV_100}. Looking at the variation of the effective mass with $M$, we can see that the change in $|m_{ee}^{(4)}|$ can be quite dramatic for very small values of $M$: while the keV neutrino mass only changes by a factor of $1.6$ from the upper left to the lower left panel in Fig.~\ref{fig:keV_100}, we can see that for $M = 1$~keV the region where a full cancellation of the effective mass is possible is very different from the one for $M = 1.6$~keV. This is a characteristic behaviour of the elements of the neutrino mass matrix~\cite{Lindner:2005kr,Merle:2006du}: the different contributions to the effective mass can be viewed as vectors in the complex plane which can, depending on their respective lengths and the values of their phases, add up to zero or not. Similarly in our case, if the lengths are appropriate, a cancellation can be achieved by varying the Majorana phases $\alpha$, $\beta$, and $\gamma$ accordingly. 
However, if the lengths of the vectors do not have the correct proportions (e.g., if one of the four is considerably longer than the three other ones), one can never build a zero length vector out of them, and then the effective mass will always be non-zero. From the plots we see that for $M = 1$~keV the best-fit regions and the 3$\sigma$ regions are both different compared to the standard case. For $M=1.6$~keV, instead, mainly the best-fit regions change compared to the standard case.

We can try to understand this behaviour analytically by glancing at Eqs.~\eqref{eq:mee_1} and~\eqref{eq:mee_2}. Using the OLD limit, the keV neutrino contribution $M \theta^2$ is $\lesssim 0.0055$~eV ($\lesssim 0.0005$~eV) for $M=1$~keV ($M=1.6$~keV). Since the decisive region in the plot is the area around $m \simeq 0$ for NO, we can approximate the situation by normal hierarchy: $m_1 \simeq 0$, $m_2 \simeq \sqrt{\Delta m^2_\odot}$, $m_3 \simeq \sqrt{\Delta m^2_A}$. According to Eq.~\eqref{eq:mee_1}, this leads to
\begin{equation}
 |m_{ee}^{(3)}| \simeq |\sqrt{\Delta m^2_\odot} s_{12}^2 c_{13}^2 e^{2i \alpha} + \sqrt{\Delta m^2_A} s_{13}^2 e^{2i \beta}|  \simeq |\sqrt{\Delta m^2_\odot} s_{12}^2 + \sqrt{\Delta m^2_A} s_{13}^2 e^{2i (\beta- \alpha)}|.
 \label{eq:an_1}
\end{equation}
Varying the phases for the best-fit values of the OLD oscillation parameters (the phases and the OLD oscillation parameters within their $3\sigma$ ranges), this quantity is between $0.0023$ and $0.0033$~eV (between $0$ and $0.0059$~eV). Clearly, both the best-fit value and the $3\sigma$ can be of the order of the keV neutrino contribution for $M=1$~keV, and thus a cancellation to zero is possible. The $M=1.6$~keV contribution, in turn, is always smaller for the best-fit case, which makes a cancellation to practically zero impossible. However, if the oscillation parameters are varied within their $3\sigma$ ranges, then a cancellation can indeed appear for $M=1.6$~keV, cf.\ first column of Fig.~\ref{fig:keV_100}. Since the plots corresponding to low values of $M$ differ significantly from the standard plots, cf.\ Fig.~\ref{fig:mee}, this yields an interesting connection between production mechanisms, which allow for lower keV sterile neutrino masses, and experiments on neutrino-less double beta decay. 

Let us now see how this situation has changed after the new limits from Chandra. For the NEW conservative bound, the main difference with respect to the standard contribution is present for $M$=1~keV. In this case, the best-fit and 3$\sigma$ regions for both NO and IO are different compared to the standard effective mass. On the other hand the new bounds, even if taken into account in a conservative way, are so strong that the effect of a sterile neutrino with $M$=1.6~keV is already really tiny. For the case of the NEW restrictive bound, practically no effect is visible at all. This result shows that the actual X-ray bounds taken at face value are extremely strong, and they suppress any possible influence of the keV neutrino on the effective mass, even for a mass as low as $M=1$~keV. 

If, on the other hand, the DW-bound is taken at face value, any influence is wiped out even for the OLD scenario, as can be seen by comparing the red lines in Fig.~\ref{fig:keV_100} with the general allowed regions presented earlier in Fig.~\ref{fig:mee}. This is perfectly consistent with the solid blue line in the left panel of Fig.~\ref{fig:Bounds_sterile}. Any possible contribution to $0\nu\beta\beta$ is reduced to negligible values.

\begin{figure}[t!]
\hspace{-1.0cm}
\begin{tabular}{lcr}
\includegraphics[width=5.5cm]{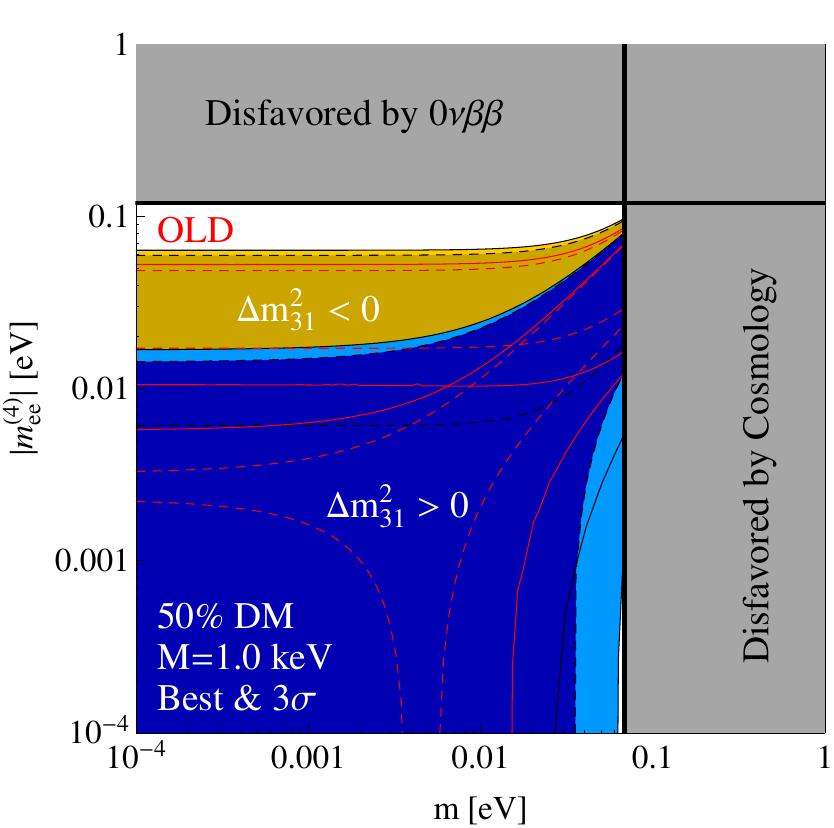} & \includegraphics[width=5.5cm]{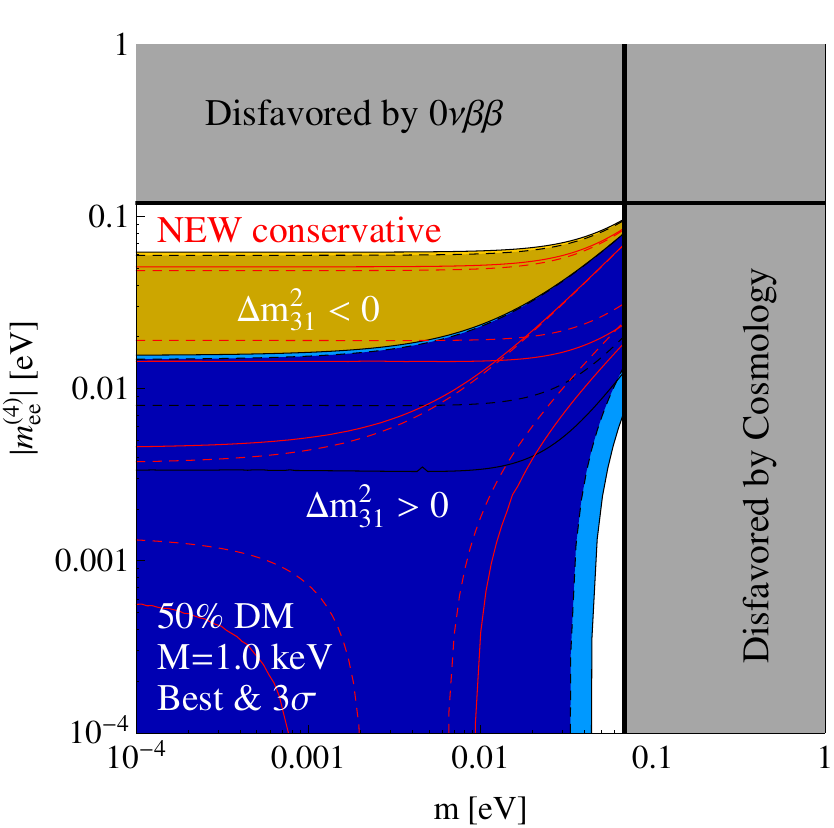} & \includegraphics[width=5.5cm]{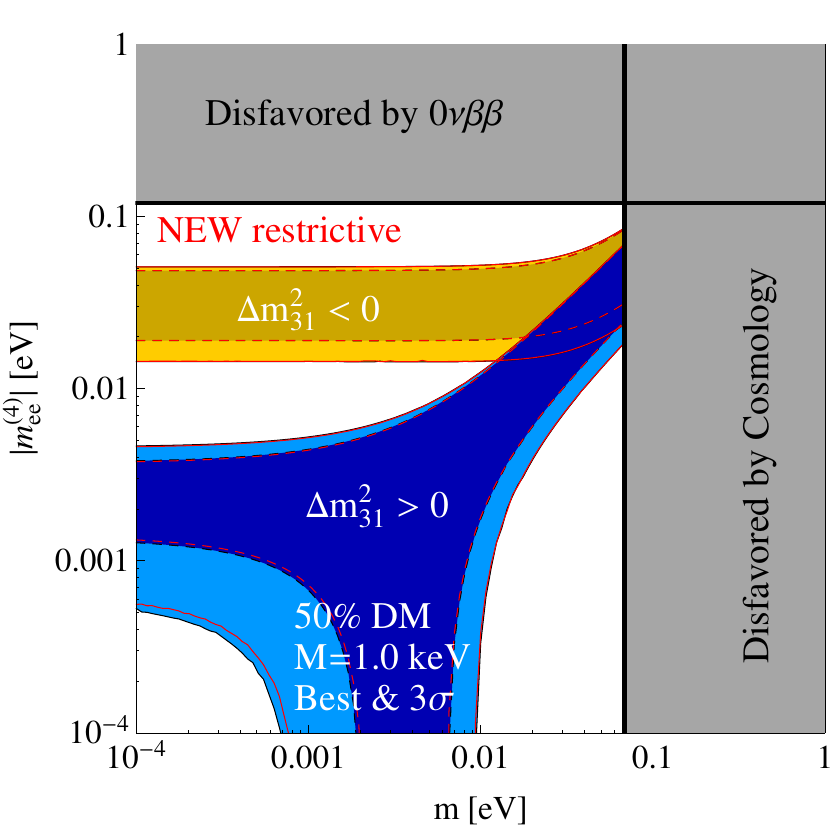}\\
\includegraphics[width=5.5cm]{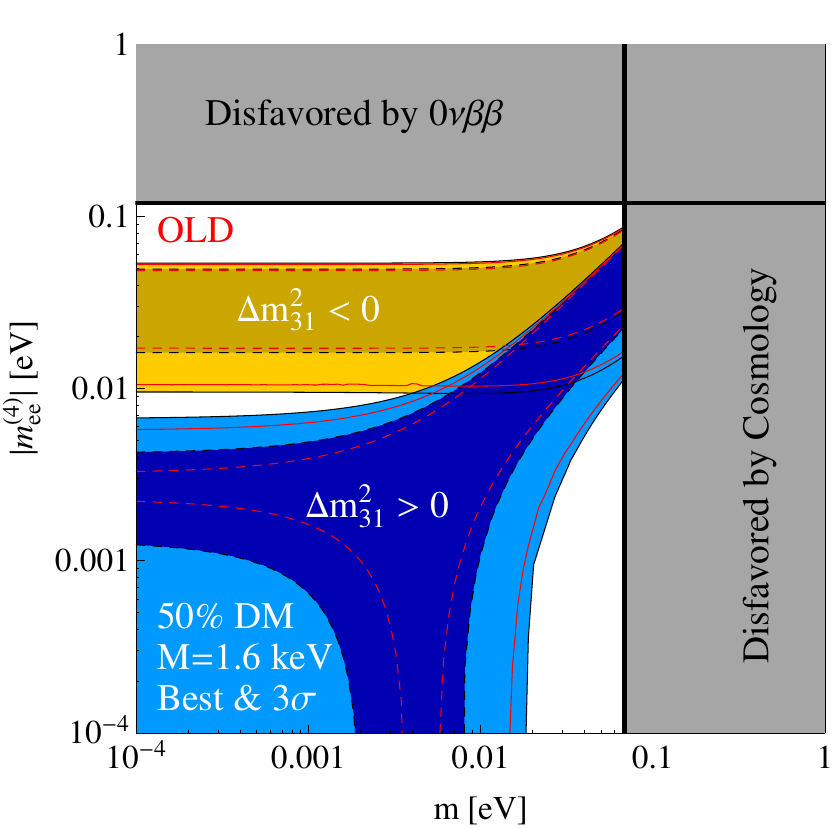} & \includegraphics[width=5.5cm]{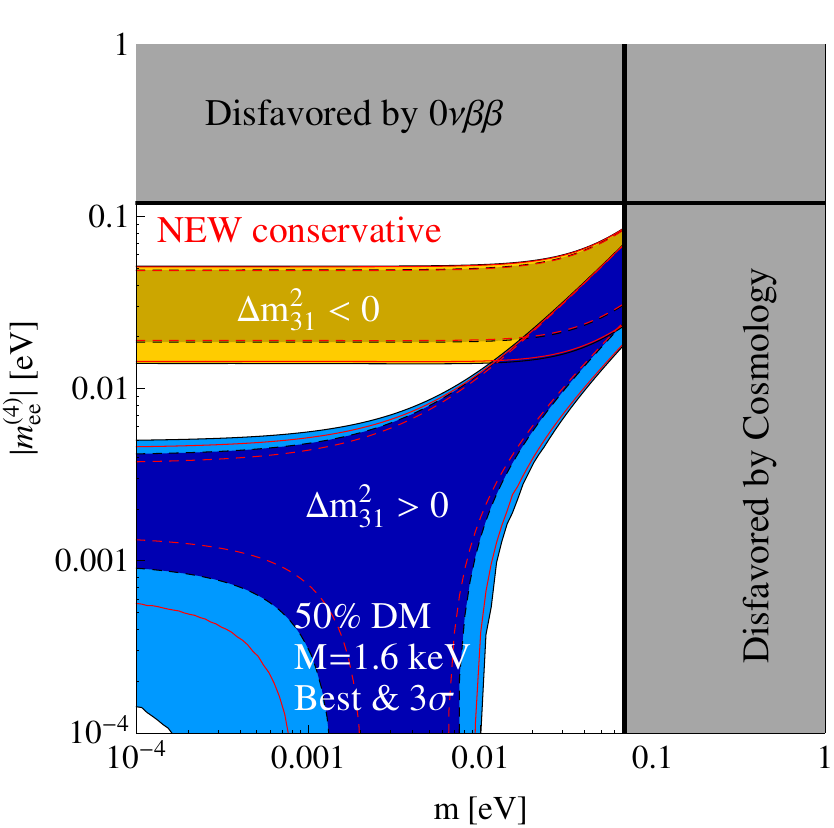} & \includegraphics[width=5.5cm]{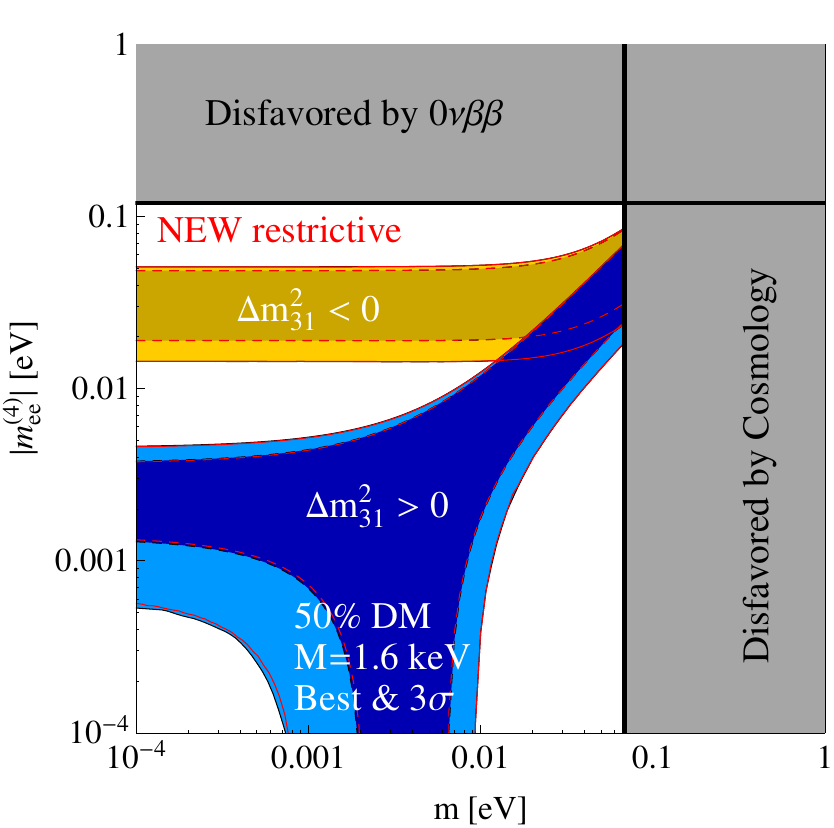}
\end{tabular}
\caption{\label{fig:keV_50} Same as Fig.~\ref{fig:keV_100}, but assuming that the sterile neutrinos constitute only $50\%$ of the DM present in the Universe.}
\end{figure}

In Fig.~\ref{fig:keV_50}, we present the results for the case in which the sterile neutrino constitutes only 50\% of the DM. This could be the case in a scenario with more than one type of DM, see e.g.\ Ref.~\cite{Heeck:2012bz} for a setting where keV DM is mixed with heavier DM. The rough effect of this assumption is that the upper bound on the active-sterile mixing $\theta^2$ gets weaker by a factor of two, simply because the amount of keV sterile neutrino DM, and hence the number of expected decays $N \to \nu \gamma$, is reduced by the same factor. Note that this is only an approximation, since the statistical error of the number of possible events in a certain bin (and most probably also some systematical errors) depend on the amount of keV sterile neutrinos in the Universe. However, such effects should be small compared to the errors involved in the nuclear physics uncertainties in $0\nu\beta\beta$~\cite{Rodejohann:2011mu}. On the other hand, since in this case we have less DM made of keV neutrinos, the 
bound arising from trying to avoid producing too much DM by the DW-mechanism does in fact become \emph{stronger}. This can, again, be seen from the red lines in the plots.

For the situation valid before 2011, a visible effect could have been present for both example masses of $1$ and $1.6$~keV in the general case (i.e., disregarding the DW-bound). Note that, while the keV neutrino mass only changes by a factor of $1.6$ from the upper left to the lower left Fig.~\ref{fig:keV_100}, we can see that, for $M = 1$~keV, the region where a full cancellation of the effective mass is possible is very big for NO in case the best-fit of the oscillation parameters are taken or in case they are varied within the $3\sigma$ allowed ranges. However, for $M = 1.6$~keV the cancellation is possible for NO just in case the oscillation parameters are varied within their $3\sigma$ ranges.

Again aiming at analytically understanding this behaviour, for the OLD limit, 
the contribution of keV neutrino making up only $50\%$ of all DM in the Universe amounts to $\lesssim 0.0110$~eV ($\lesssim 0.0009$~eV) for $M=1$~keV ($M=1.6$~keV). The variation of $|m_{ee}^{(3)}|$ for NO is the same as described after Eq.~\eqref{eq:an_1}. Evidently, for $M=1$~keV a cancellation is possible for NO in case the variation is both within the best-fit and the $3\sigma$ ranges. For $M=1.6$~keV, instead, the cancellation appears just in case the neutrino oscillation parameters are changed within their $3\sigma$ ranges. These considerations are in full accordance with the plots, cf.\ first and second plot in the first column of Fig.~\ref{fig:keV_50}.

For the NEW conservative bound, analogously, the main differences with respect to the standard case are present for $M=1$~keV. In that case, the IO and NO regions are significantly different from that of the standard case, cf.\ right panel of Fig.~\ref{fig:mee}. For $M=1.6$~keV, instead, only a small difference is present for the best-fit and 3$\sigma$ variation of the NO case. For the NEW restrictive bound, also in this case there are no significant differences present compared to the standard light neutrino contribution.

Adding the DW-bound, if applicable, would again destroy any influence (cf.\ red lines in Fig.~\ref{fig:keV_50} and right panel of Fig.~\ref{fig:Bounds_sterile}). This illustrates that an interpretation of a possible future signal of $0\nu\beta\beta$ outside of the range predicted in Fig.~\ref{fig:mee} in terms of keV neutrinos would require a careful investigation of different production mechanisms.

\section{\label{sec:applic}On the applicability of our study}

We have studied in the previous sections the possibility that a really light keV sterile neutrino, $M \lesssim 2$~keV, could contribute to the $0\nu\beta\beta$ effective mass. In this low keV sterile neutrino mass region the DW-mechanism -- where sterile neutrinos are produced by the plasma through admixtures to active neutrinos -- would lead to a significant overabundance of DM for active-sterile mixing angles close to the upper observational X-ray bound. This seems like a tough problem, since it exactly affects the region where a potentially observable influence of the keV sterile neutrino on $0\nu\beta\beta$ could in general be possible, as illustrated by the red lines in the plots. Even in the case where keV sterile neutrinos make up only 50$\%$ of the DM in the Universe, while the X-ray bound is weakened by roughly a factor of $2$, the bound from the DW-contribution is strengthened by the same factor, since now overall less DM is present in keV steriles. However, there are different possibilities and 
different models to weaken or even evade this problem.

Another critical point of our phenomenological analysis is that we take the mass $M$ of the keV sterile neutrino to be completely independent of the lightest neutrino mass $m = m_1$ (normal ordering) or $m = m_3$ (inverted ordering). However, if the mass of the lightest active neutrino is calculated in the type~I seesaw approximation, which is possible in models with keV sterile neutrinos as long as the X-ray bound is respected~\cite{Merle:2012xq}, then these masses are related to the square of the active-sterile mixing angle, $m \simeq \theta^2 M$. Since $\theta$ and $M$ are restricted in our plots, this strict dependence would not allow to vary $m$ considerably. However, also this situation can be overcome.

We will now discuss how a model, i.e., a combination of a mass generating mechanism for light neutrinos and a production mechanism for keV sterile neutrino DM, needs to be designed in order to avoid both these restrictions and have a potentially observable effect in $0\nu\beta\beta$. Let us start with the mass generation mechanism for the light neutrinos. While a type~I seesaw mechanism~\cite{Minkowski:1977sc,Yanagida:1979as,GellMann:1980vs,Glashow:1979nm,Mohapatra:1979ia} would be strongly constrained, already a type~II~\cite{Magg:1980ut,Lazarides:1980nt} contribution -- generated by the unconstrained Yukawa coupling of the left-handed lepton doublet to a triplet scalar which obtains a vacuum expectation value -- would easily be enough to avoid the strong relation between the masses of the lightest active neutrino and of the keV neutrino. Similarly a type~III seesaw~\cite{Foot:1988aq}, where typically three fermion triplets are introduced, would considerably disentangle the relation. Even more complicated 
situations could arise in, e.g., models where the light neutrino mass is generated only at 
loop-level~(see Refs.~\cite{Ma:2006km,Zee:1985id,Babu:1988ki,Aoki:2008av,Gustafsson:2012vj} for generic examples).

What about the production mechanism? The problem is that it is not easy to overcome the DW-contribution as long as a significant active-sterile neutrino mixing is present. On the other hand, the DW-mechanism can be modified. In fact, the data tell us that it even must be modified, since it is not in agreement with all bounds if it is the sole production mechanism of keV sterile neutrino DM~\cite{Canetti:2012vf,Canetti:2012kh}. A generic method is to simply dilute an overabundance of DM in keV sterile neutrinos by the production of a significant amount of entropy due to particle decays. For example, one could use the decays of the heavier sterile neutrinos $N_{2,3}$, which could already have an effect within the $\nu$MSM and which could be strongly enhanced in the presence of further new physics~\cite{Asaka:2006ek}. This method has also been used in scenarios where the sterile neutrinos are sterile only with respect to the SM, but charged under an extended gauge group. In this way, they are thermally 
overproduced but their abundance is diluted later on by the entropy-producing decays~\cite{Bezrukov:2009th,Nemevsek:2012cd}. As shown in Fig.~\ref{fig:Bounds_sterile}, the required dilution factor would need to be of $\mathcal{O}(100)$ to generate potentially visible effects of low-mass keV neutrinos on $0\nu\beta\beta$. While such numbers are not entirely easy to achieve, there are cases known in the literature where they are possible~\cite{Ma:2012if,Bezrukov:2009th}. In such situations, even if the naive abundance calculation within the DW-mechanism was not sufficient, the more model-independent X-ray bound would nevertheless apply. However, we want to stress that, typically, one has to extend the particle content of the model compared to the $\nu$MSM in order to generate large amounts of entropy dilution and/or production~\cite{Ma:2012if,Bezrukov:2009th}.

The lesson to learn is that one has to be careful in the selection of a setting where the most general analysis actually applies. If it does, there can indeed be observable modifications of the $0\nu\beta\beta$ rate compared to the standard case of light neutrino exchange. In a time where we could expect a potential detection of the process in the near future, we should be aware of this possibility when aiming to correctly interpret upcoming experimental results. However, it is also clear that our results are not applicable for every model, and that the corresponding window in the parameter space it not very big. Depending on how strong the different observational bounds -- in particular the X-ray bound on the keV neutrino decay rate and the lower bound on the keV neutrino mass from the measurement of the Ly-$\alpha$ forest -- will get, we could close even this gap very soon.

\section{\label{sec:conc}Conclusions}

We have re-analysed the contribution of one keV sterile neutrino to neutrino-less double beta decay, focussing particularly on the low mass region. We have shown that, considering the existing X-ray limits before 2011, depending on the light neutrino mass generation and DM production mechanisms, sterile neutrinos could have had a visible influence on the effective mass if they were as light as $M<$~2~keV (had one assumed an active-sterile mixing angle close to the by then most recent observational upper bound). This was to some extent missed by earlier references. We have then updated our study considering the recent Chandra and XMM-Newton results, taking the new limits with both a conservative and a restrictive approach. We found that for the conservative case the sterile neutrinos could still produce a visible modification on the effective mass for really light mass values, $M\simeq 1$~keV, while for higher masses the effect is tiny. However, when using the Chandra and XMM-Newton limits at face value, the 
influence of a keV sterile neutrino on the effective mass is completely washed out. This is a consequence of the really strong new limits on the active-sterile neutrino mixing angle that are present at the moment from the satellite experiments mentioned. Furthermore, if a significant amount of the DM is produced by the DW-mechanism, this imposes an upper bound on the active-sterile mixing angle which can be even stronger than the X-ray bound and would completely destroy any possible effect. This could only be avoided if a significant amount of entropy is produced in a certain decay.

We have then moved into analysing the case in which the sterile neutrino constitutes only 50\% of the DM. Also in that case, with the limits before 2011 taken at face value, the effect on the effective mass could actually be significant for light sterile neutrinos, $M<$~2~keV. For the conservative scenario, the biggest effect is present close to the lower bound, $M\simeq 1$~keV, but a small modification is also present for $M\simeq 1.6$~keV. Considering the restrictive limit, no effect is visible even for really light sterile neutrino masses. Furthermore, in this case, the DW-bound gets even stronger (and thus more destructive), since even less DM is allowed to be produced by the DW-mechanism.

In addition, we have discussed in which types of settings and models visible effects on the $0\nu\beta\beta$ rate could appear. While certain settings are very tightly constrained, others could at least potentially allow for an observable effect. Even though the effect might only be present in some settings, we should nevertheless be aware of this possibility, since it could potentially become interesting in case soon-to-be-expected experimental results on $0\nu\beta\beta$ would lead to something unexpected.

\section*{\label{sec:ack}Acknowledgements}

We would like to thank J.~Barry, J.~Heeck, A.~Kusenko, M.~Loewenstein, W.~Rodejohann, and O.~Ruchayskiy for useful discussions, and we are particularly grateful to J.~Bergstr\"om and G.~Pedaletti for help concerning the data handling. AM acknowledges support by a Marie Curie Intra-European Fellowship within the 7th European Community Framework Programme FP7-PEOPLE-2011-IEF, contract PIEF-GA-2011-297557. VN acknowledges support by the Spanish research Grant FPA2010-20807 and by the consolider-ingenio 2010 program grants CUP (CSD-2008-00037), and by the CPAN. Furthermore, AM and VN both acknowledge partial support from the European Union FP7 ITN-INVISIBLES (Marie Curie Actions, PITN-GA-2011-289442).

\bibliographystyle{./apsrev}
\bibliography{./keV_0nbb}

\end{document}